\documentclass[twocolumn,showpacs,pre,aps]{revtex4}

\usepackage{dcolumn}
\usepackage{bm}

\begin{document}

\title{Wave Functions for Open Quantum Systems \\
and Stochastic Schr\"{o}dinger Equations}

\author{Yuriy E. Kuzovlev}
\email{kuzovlev@kinetic.ac.donetsk.ua}
\affiliation{A.A.Galkin Physics and Technology Institute
of NASU, 83114 Donetsk, Ukraine}

\date{\today}

\begin{abstract}
It is shown that evolution of an open quantum system can be exactly
described in terms of wave function which obeys Schr\"{o}dinger
equation with randomly varying parameters whose statistics is
universally determined by separate dynamics of the system's
environment. Corresponding stochastic evolution of the wave function
is unitary on average, and this property implies optical theorem for
inelastic scattering as demonstrated by the example of
one-dimensional conducting channel with thermally fluctuating
potential perturbation.
\end{abstract}

\pacs{02.50.-r, 05.40.-a, 05.40,.Jc, 31.15.Kb}

\maketitle

{\,\bf I.\,} Generally, even simple quantum systems have no definite
wave function and require the density matrix language instead
\cite{ll}. But if the density matrix, $\,R(t)\,$, evolves under von
Neumann equation, still one can write
\begin{equation}
\begin{array}{c}
R(t)=\,\sum_{\alpha}\Psi_{\alpha}(t)\, P_{\alpha}\,\Psi_{\alpha
}^{\dagger}(t)\,\,,\label{expan}
\end{array}
\end{equation}
with $\,\Psi_{\alpha}(t)\,$ obeying the Schr\"{o}dinger equation, and
say that the system has random wave function whose randomness is
determined by initial conditions only. This is not the case as soon
as the system becomes open, i.e. constantly interacts with the rest
of the nature, and therefore is governed by some kinetic equation.
However, in the framework of the ``stochastic representation'' of
quantum interactions \cite{i1,i2,i3}, or deterministic interactions
at all \cite{i3,i4}, we can again rehabilitate wave function if say
that it is stochastic one being governed by a well-defined stochastic
Schr\"{o}dinger equation. Let us consider this potential convenience
more attentively.

{\bf\,II.\,}  Suppose that ~{\bf(i)}~ a composite quantum system
``D+B'', with ``D'' being a dynamic subsystem under direct interest
and ``B'' its environment (``thermal bath'' or `all the Universe'' or
some other), has the bilinear Hamiltonian
\begin{equation}
\begin{array}{c}
H(t)=\,H_d(t)+H_b+B_j\,D_j\,\,\,\label{ham}
\end{array}
\end{equation}
(repeated indices imply summation over them), where operators
$\,H_d\,,D_j\,$ relate to ``D'' while $\,H_b\,,B_j\,$ to ``B'';
~{\bf(ii)}~ common density matrix of ``D+B'', $\,\rho(t)\,$, obeys
the von Neumann equation, $\,d\rho(t)/dt=i[\rho(t),H(t)]/\hbar\,$,
and ~{\bf(iii)}~ in the old days ``D'' was statistically independent
of ``B'', $\rho(t_0)=\rho_{D}(t_0)\times\rho_{B}(t_0)\,$ (where e.g.
$\,t_0\rightarrow -\infty\,$).

Then marginal density matrix of the subsystem  ``D'',
$\,\rho_D(t)\equiv $~Tr$_B\,\rho(t)\,$, can be represented
\cite{i1,i2} as statistical average,
\begin{equation}
\begin{array}{c}
\rho_D(t)\,=\left\langle R(t)\right\rangle\,\,\,,\,\label{pdm}
\end{array}
\end{equation}
of a random density matrix, $\,R(t)\,$, which satisfies {\it the
stochastic von Neumann equation}
\begin{equation}
\frac {dR}{dt}=\frac i{\hbar}\left\{[R,H_d(t)]+\xi^{*}_j(t)R
D_j-\xi_j(t)D_jR\,\right\}\,\,\,,\label{se}
\end{equation}
where $\,\xi_j(t)\,$ are complex random processes which {\it commute
one with another and with any other objects}, the superscript $\,*\,$
means complex conjugation and $\,\left\langle...\right\rangle\,$
statistical average with respect to $\,\xi_j(t)$. Sometimes it is
convenient to write
\[
\begin{array}{c}
\xi(t)=x(t)+i\hbar\, y(t)/2\,\,,\,\,\,\xi^{*}(t)=x(t)-i\hbar\,
y(t)/2\,\,,
\end{array}
\]
with $\,x_j(t)\,$ and $\,y_j(t)\,$ treated as real-valued random
processes. Statistics of $\,\xi(t)$'s is completely determined by
separate dynamics of the subsystem ``B''~:
\begin{equation}
\begin{array}{c}
\left\langle \,\,\exp \{\,\int [\,a^{*}_j(t)\xi _j(t)+a_j(t)
\xi^{*} _j(t)\,]\,dt\}\,\right\rangle =\label{ncf}
\end{array}
\end{equation}
\[
=\text{Tr}_B\,\,\overrightarrow{\exp }\left[ \int
a_j(t)B_j(t)dt\right]\overleftarrow{\exp }\left[ \int
a^{*}_j(t)B_j(t)dt\right]\rho _{b0}\,\,,
\]
with $\,B_j(t)\equiv\exp[iH_bt/\hbar]\,B_j\exp [-iH_bt/\hbar]\,$,
left arrow and right arrow denoting chronological and
anti-chronological time ordering, respectively, $\,a_j(t)\,$ and
$\,a^{*}_j(t)\,$ being arbitrary test functions, or probe functions
(which are not necessarily mutually conjugate), and
~$\,\rho_{b0}\equiv \,\exp[iH_bt_0/\hbar]\,\rho_B(t_0)\exp
[-iH_bt_0/\hbar]\,$.

Clearly, any solution to Eq.\ref{se} can be represented in the form
(\ref{expan}), where wave functions (quantum states)
$\Psi_{\alpha}(t)\,$ satisfy the {\it stochastic Schr\"{o}dinger
equation}
\begin{equation}
\frac {d}{dt}\,\Psi(t) =-\frac i{\hbar}\{H_d(t)+\xi_j(t)
D_j+\epsilon(t)\}\,\Psi(t)\,\,\label{sse}
\end{equation}
with initial condition ~$\Psi_{\alpha}(t_0)\,
P_{\alpha}\,\Psi_{\alpha }^{\dagger}(t_0)=\rho_D(t_0)\,$, and
$\,\epsilon(t)\,$ is a suitable (real-valued) gauge function. At
that, formula (\ref{ncf}) supplies exhaustive statistical information
about the noises $\,\xi_j(t)$. To avoid a need for $\,\epsilon(t)\,$,
of course, it is reasonable to assume
\begin{equation}
\begin{array}{c}
\text{Tr}_D\,D_j=0\,\,,\,\,\,\,\text{Tr}_B\,B_j=0\,\,,\label{zt}
\end{array}
\end{equation}
thus excluding from (\ref{ham}) trivial contributions without factual
interaction.

{\bf\,III.\,}~Because of the complexity of $\,\xi_j(t)$, the
stochastic evolution under Eq.\ref{sse} is not unitary:
\begin{equation}
\frac {d}{dt}\,\langle\Psi_{\alpha}(t)|\Psi_{\beta}(t)\rangle =
y_j(t)\langle\Psi_{\alpha}(t)|D_j|\Psi_{\beta}(t)\rangle\neq 0\,\
\label{nu}
\end{equation}
But this evolution is {\it unitary on average}~:
\begin{eqnarray}
\frac {d}{dt}\,\left\langle\,
\langle\Psi_{\alpha}(t)|\Psi_{\beta}(t)\rangle\,\right\rangle
=\langle
y_j(t)\langle\Psi_{\alpha}(t)|D_j|\Psi_{\beta}(t)\rangle\,\rangle
= 0\,\,, \label{au}\\
\left\langle\,\,\langle\Psi_{\alpha}(t)|\Psi_{\beta}(t)\rangle\,
\,\right\rangle =
\langle\Psi_{\alpha}(t_0)|\Psi_{\beta}(t_0)\rangle\propto
\delta_{\alpha\beta}\,\,\,\,\,\,\,\,\,\ \label{au1}
\end{eqnarray}
The statement (\ref{au1}) follows from (\ref{au}) and
$\,\Psi_{\alpha}(t_0)$'s definition, while the statement (\ref{au})
from remarkably singular statistical properties of $\,y_j(t)\,$
\cite{i1,i2,i3,i4,i5}.

To see the $\,y_j(t)$'s singularity, it is sufficient to assign in
(\ref{ncf}) $\,a(t)=iu(t)/\hbar\,$ and $\,a^{*}(t)=-iu(t)/\hbar\,$
when $\,t>\theta\,$, with real test functions $\,u(t)\,$ and
$\,\theta>t_0$. Then ~$\,\overleftarrow{\exp }\left[ \int
a^{*}(t)B(t)dt\right]=U\,\overleftarrow{\exp }\left[ \int^{\theta}
a^{*}(t)B(t)dt\right]\,$ and ~$\,\overrightarrow{\exp }\left[ \int
a(t)B(t)dt\right]=\overrightarrow{\exp }\left[ \int^{\theta}
a(t)B(t)dt\right]U^{\dagger}\,\,$, where $\,U=\overleftarrow{\exp
}\left[ -i\int_{\theta} u(t)B(t)dt/\hbar\right]\,$. Since $\,U\,$ is
unitary, $\,U^{\dagger}U=1\,$, in fact it disappears from r.h.s. of
(\ref{ncf}):
\begin{equation}
\begin{array}{c}
\langle \,\exp \{\,\int_{\theta}\,u(t)\,y(t)\,dt\}\exp
\{\,\int^{\theta}[\,a^{*}(t)\xi(t)+a(t)\xi^{*}
(t)\,]\,dt\}\rangle \\
=\text{Tr}_B\,\,\overrightarrow{\exp }\left[ \int^{\theta}
a(t)B(t)dt\right]\overleftarrow{\exp }\left[ \int^{\theta}
a^{*}(t)B(t)dt\right]\rho _{b0}\,
\end{array} \label{ncf1}
\end{equation}
Here from it follows that any statistical correlation between
$\,y(t)\,$ and any earlier variables is zero. In particular, the
equality (\ref{au}) is valid because $\,\Psi(t)$'s undergo the
causality principle and depend on earlier $\,\xi(<t)$'s only. The
Eq.\ref{ncf1} implies also ~$\,\langle y(t_1)...y(t_n)\rangle =0\,$~,
~~$\,\langle y(t)x(t^{\prime}<t)\rangle =0\,$, etc.

At the same time, generally $\,\langle x(t)y(t^{\prime}<t)\rangle
\neq 0\,$ and other $\,y(t)$'s correlations with later variables also
are nonzero.

Let the noise is stationary, that is $\,[H_b,\rho_{b0}]=0\,$, and
besides has zero mean value, $\,\langle \xi_j(t)\rangle =0\,$. The
latter is ensured by the natural condition
$\,\text{Tr}_B\,B_j\,\rho_{b0}\,=0\,$. Then \cite{i1}, according to
Eq.\ref{ncf},
\begin{equation}
\begin{array}{c}
\langle \xi^{*}_j(\tau )\xi _m(0)\rangle =K_{jm}(\tau )\,\,,\label{cf}\\
K_{jm}(\tau )\equiv \text{Tr}_B\,\,B_j(\tau )B_m(0)\,\rho_{b0}\,
=K_{mj}^{*}(-\tau)\,\,,\\
\langle \xi _j(\tau )\xi _m(0)\rangle =\langle \xi^{*}_j(\tau
)\xi^{*}_m(0)\rangle ^{*}=\\
=K_{jm}(\tau )\theta(\tau)+K_{jm}^{*}(\tau )\theta(-\tau)\,\,,
\end{array}
\end{equation}
where $\,\theta(\tau)\,$ is Heaviside step function. Consequently,
\begin{equation}
\begin{array}{c}
K_{jm}^{xx}(\tau )\equiv \langle x_j(\tau )x_m(0)\rangle =\text{Re}
\,K_{jm}(\tau )\,\,,\label{rcf}\\
K_{jm}^{xy}(\tau )\equiv \langle x_j(\tau )y_m(0)\rangle
=(2/\hbar)\,\theta(\tau)\,\text{Im}\,K_{jm}(\tau )
\end{array}
\end{equation}

If $\,\rho_{b0}\,$ is canonical distribution, $\,\rho _{b0}\propto
\exp (-H_b/T)\,$, then the identity $\,\rho
_{b0}B(t)\rho^{-1}_{b0}=B(t+i\hbar /T)\,$ takes place and produces
additional symmetry relation
\begin{equation}
\begin{array}{c}
K_{jm}(\tau -i\hbar /2T)=K_{mj}(-\tau-i\hbar /2T)\,\,\label{T}
\end{array}
\end{equation}
The latter if combined with (\ref{cf}) results in the
``fluctuation-dissipation relations'' between spectral components of
correlators $\,K_{jm}^{xx}(\tau )\,$ and $\,K_{jm}^{xy}(\tau )\,$ as
expressed by their Fourier expansions:
\begin{equation}
\begin{array}{c}
K_{jm}^{xx}(\tau )=\int_0^\infty \cos (\omega \tau )\,\sigma
_{jm}(\omega )d\omega \,\,,\label{kxx}
\end{array}
\end{equation}
\begin{equation}
K_{jm}^{xy}(\tau )=-\frac {2\theta (\tau)}{\hbar}\int_0^\infty \sin
(\omega \tau )\tanh \left[\frac {\hbar\omega}{2T}\right] \sigma
_{jm}(\omega )d\omega\,\,,\label{kxy}
\end{equation}
with $\,\sigma _{jm}(\omega )\,$ being some non-negatively defined
spectrum matrix. Unfortunately, in \cite{i1} formula (\ref{kxy}) was
written with wrong plus sign on its r.h.s., and then this mistake
migrated to \cite{i2,i3,i4}. But, fortunately, it could not influence
results of applications of (\ref{kxy}) since all that were even
functions of $\,K_{jm}^{xy}(\tau )\,$. The term
``fluctuation-dissipation'' reflects that $\,K_{jm}^{xy}(\tau )\,$
and other cross correlations between $\,x(t)$'s and $\,y(t)$'s
describe also response of ``B'' to its perturbation by ``D'' and thus
possible energy outflow from ``D'' into ``B''. For details and
extensions of the stochastic representation (\ref{pdm})-(\ref{sse})
see \cite{i1,i2,i3,i4,i5}.

{\bf\,IV.\,}~For illustration, let ``D'' be one-dimensional
conduction channel with finite energy band, that is formed by
discrete sites $\,s=...-1,0,1,...\,$, and its perturbation by ``B''
acts as randomly varying potential localized at the only site
$\,s=0\,$. Thus $\,\xi_j(t)D_j\equiv \xi (t)D\,$ where
$\,D_{ss^{\prime}}=1\,$ if $\,s=s^{\prime}=0\,$ and
$\,D_{ss^{\prime}}=0\,$ otherwise. Definition of the own Hamiltonian
of ``D'' is convenient in the momentum space:
$\,\{H_d(t)\}_{kk^{\prime}}=\delta (k-k^{\prime})E(k)\,$, where
$\,-\pi<k<\pi\,$ and, for instance, $\,E(k)=\Delta E (1-\cos\,k)/2\,$
with $\,\Delta E\,$ being channel's bandwidth .

In this example, it is natural to treat $\,\Psi(t)\,$ in the
coordinate space, choose $\,\epsilon (t)=0\,$, and transform
Eq.\ref{sse} into integral equation:
\begin{equation}
\begin{array}{c}
\Psi(t,s)=\Psi^{0}(t,s)+\Psi^S(t,s)\,\,,\\
\Psi^S(t,s)=-\frac
i{\hbar}\int_{t_0}^{t}G(t-t^{\prime},s)\xi(t^{\prime})
\Psi(t^{\prime},0)\,dt^{\prime}\,\, \label{ise}
\end{array}
\end{equation}
where $\,\Psi^{0}(t,s)\,$ and $\,\Psi^S(t,s)\,$ are free and
scattered parts of the wave function, respectively, and
\begin{equation}
\begin{array}{c}
G(\tau,s)=\int_{-\pi}^{\pi}\exp \left[iks-iE(k)\tau/\hbar
\right]\,dk/2\pi \label{g00}
\end{array}
\end{equation}
is free Green function.

Consider stochastic scattering of a single particle. Let initially,
at time $\,t_0=-\tau_0 <0\,$, it had wave number $\,k_0>0\,$ and
velocity $\,v_0=dE(k_0)/\hbar dk_0\,$, and its wave packet took place
$\,-L<s<0\,$ with $\,L\approx v_0\tau_0
>>2\pi /k_0\,$. Then we observe the particle at time $\,t>0\,$.
At that, we are interested in the case when duration of the
scattering process, $\,\sim L/v_0\,$, essentially exceeds correlation
time, $\,\tau_c\,$, of the scatterer fluctuations $\,\xi(t)$. This is
possible if $\,L\,$ is sufficiently large while $\,v_0\,$
sufficiently small, at least if $\,L>L_c=\sqrt{\pi \Delta
E\tau_c/\hbar}\,$ (for example, if $\,\Delta E=1\,$eV and each site
occupies $\,3\cdot 10^{-8}\,$cm then $\,L_c\sim \sqrt{\tau_c}\,$
where $\,\tau_c\,$ is expressed in seconds and $\,L_c$ in
centimeters).

Obviously, now the equality (\ref{au1}) can be written as
\begin{equation}
\begin{array}{c}
\langle\sum |\Psi(t,s)|^2\rangle=\sum |\Psi (t_0,s)|^2=
\,\,\,\,\,\,\,\,\,\,\,\,\,\,\,\\
\,\,\,\,\,\,\,\,\,\,\,\,\,\,\,\,\,=\sum |\Psi^{0}(t_0,s)|^2=\sum
|\Psi^{0}(t,s)|^2\, \label{au2}
\end{array}
\end{equation}
Combining this with the first of Eqs.\ref{ise} one easy obtains
\begin{equation}
\begin{array}{c}
\sum\, [\,\Psi^{0*}\langle\Psi^S\rangle
+\Psi^0\langle\Psi^{S*}\rangle\,] + \langle\,\sum |\Psi^S|^2\,\rangle
=0\,\,, \label{au3}
\end{array}
\end{equation}
where all functions are related to the observation time $\,t\,$. This
formula is nothing but ``{\it the averaged optical theorem for
inelastic scattering}'' (clearly, fluctuations $\,\xi(t)\,$ in the
course of scattering can make it inelastic).

As it was agreed, at time $\,t>0\,$ the scattering event finishes,
therefore, we can express final quantum probabilities of the
particle's reflection, $\,\Re\,$, and transmission,
$\,\mathcal{T}\,$, simply as
\begin{equation}
\begin{array}{c}
\Re =\sum_{s<0}|\Psi |^2\, = \sum_{s<0}|\Psi^S|^2\,= \frac 12 \sum
|\Psi^S|^2\,\,,\\\mathcal{T}=\sum_{s>0}|\Psi |^2\,\,\,\,\,\,\,\,\,
\label{re}
\end{array}
\end{equation}
We took into account also mirror symmetry of the scattering: the
Green function $\,G(t,s)\,$ and thus $\,\Psi^S(t,s)\,$ are even
functions of $\,s\,$ (at least when $\,E(k)\,$ is an even function).
Now, taking average of (\ref{re}) and applying (\ref{au2}) and
(\ref{au3}), we obtain
\begin{equation}
\begin{array}{c}
\langle\,\Re\,\rangle = -\text{Re}\,\sum\,
\Psi^{0*}(t,s)\,\langle\Psi^S(t,s)\rangle
=1-\langle\,\mathcal{T}\,\rangle\,\label{mre}
\end{array}
\end{equation}
Hence, the optical theorem helps to find the mean reflection
probability $\,\langle\,\Re\,\rangle\,$ directly from mean value of
$\,\Psi(t,s)\,$, without knowing its second-order statistics.

Another useful expression for $\,\langle\,\Re\,\rangle\,$ arises if
combine (\ref{mre}) with (\ref{ise}) and then apply the group
property ~$\,\sum G(\tau_1,s-s^{\prime})G(\tau_2,s^{\prime})
=G(\tau_1+\tau_2,s)\,$. This yields
\begin{equation}
\begin{array}{c}
\langle\,\Re\,\rangle = \text{Re}\,\,\frac i\hbar \int_{t_0}^{t}\,
\Psi^{0*}(t^{\prime},0)\,\langle\xi(t^{\prime})
\psi(t^{\prime})\rangle \,dt^{\prime}\,\,,\label{mre1}
\end{array}
\end{equation}
where $\,\psi (t)\equiv \Psi(t,0)\,$ satisfies closed integral
equation
\begin{equation}
\begin{array}{c}
\psi(t)=\Psi^{0}(t,0)-\frac
i{\hbar}\int_{t_0}^{t}G_0(t-t^{\prime})\xi(t^{\prime})
\psi(t^{\prime})\,dt^{\prime}\,\, \label{ise1}
\end{array}
\end{equation}
with $\,G_0(\tau)\equiv G(\tau,0)\,$.

Alternatively, at $\,t>0\,$ we can consider amplitudes of outgoing
waves,
\[
\begin{array}{c}
A_k=A_k^0+A_k^S\,\,,\,\,\,\,A_k^0=\delta(k-k_0)\,\,,
\end{array}
\]
and the probabilities of corresponding particular variants of
scattering,
\begin{equation}
\begin{array}{c}
P_k=|A_k|^2\,\,,\\ \label{prob} \langle P_k\rangle =\langle
|A_k|^2\rangle =|\langle A_k\rangle |^2+\langle A_k\,,A_k^{*}\rangle
\end{array}
\end{equation}
Here and below $\,\langle a\,,\,b\rangle\equiv\langle a\,b\rangle
-\langle a\rangle\langle b\rangle\,$ is Malakhov's cumulant bracket.
If $\,\Psi^0(t,s)\,$ was normalized to unit, then the Eqs.\ref{ise}
are equivalent to
\begin{equation}
A^S_k=-\frac {i}{\hbar \sqrt{L}}\int_{t_0}^{t}\exp[i E(k)(t^{\prime
}-t_0)/\hbar]\,\,\xi(t^{\prime})\psi(t^{\prime})\,dt^{\prime}\
\label{amps}
\end{equation}
In terms of $\,A^S_k\,$, the optical theorem reads
\begin{equation}
\text{Re\,}\langle A_{k_0}^S\rangle =\text{Re\,}\langle
A_{-k_0}^S\rangle =-\langle \Re \rangle\,\,,\label{ot}
\end{equation}
where again the symmetry is taken into account.

Let us divide $\,\Re \,$ and $\,\mathcal{T} \,$ into elastic and
inelastic parts marked by superscripts ``el'' and ``in''. Obviously,
\begin{equation}
\begin{array}{c}
\langle\Re^{el}\rangle \equiv \langle P_{-k_0}\rangle \,\,,\,\,\,
\langle\mathcal{T}^{el}\rangle \equiv \langle
P_{k_0}\rangle\,\label{els}
\end{array}
\end{equation}
These identities together with Eq.\ref{prob} and Eq.\ref{ot} yield
\begin{equation}
\begin{array}{c}
\langle\Re^{in}\rangle =\langle\Re\rangle-\langle\Re^{el}\rangle \leq
\langle\Re\rangle-\langle\Re\rangle ^2\,\,,\\
\langle\mathcal{T}^{in}\rangle
=\langle\mathcal{T}\rangle-\langle\mathcal{T}^{el}\rangle \leq
\langle\mathcal{T}\rangle-\langle\mathcal{T}\rangle ^2
\,\,\,\,\label{ins}
\end{array}
\end{equation}
In fact, that are one and the same relation, because $\,\Re^{in}
=\mathcal{T}^{in}\,$ due to the mirror symmetry. Here from we obtain
the restriction of total inelastic scattering:
\begin{equation}
\begin{array}{c}
\langle\Re^{in}\rangle +\langle\mathcal{T}^{in}\rangle \leq
\,2\,\langle\Re\rangle\,[1-\langle\Re\rangle ]\,\leq 1/2
\,\label{restr}
\end{array}
\end{equation}
This is consequence of the ``unitarity on average''.

{\bf\,V.\,}~ Now consider the stochastic scattering more concretely,
basing on the Eq.\ref{ise1} and assuming that $\,\xi(t)\,$ is a
stationary random process, in particular, its mean value is
time-independent: $\,\langle\xi(t)\rangle =\langle x(t)\rangle
=\overline{x}=\,$const~ (of course, $\,\overline{x}\,$ must be real).

Let us use the star $\,\star\,\,$ as symbol of causal convolution:
~~$\,(a\star \,b)(t)\equiv $ $\int_{t>t^{\prime}}a(t-t^{\prime})
b(t^{\prime})\,dt^{\prime}\,\,$. Due to the stationarity, when
averaging Eq.\ref{ise1} we can write
\begin{equation}
\begin{array}{c}
\,\langle\xi(t)\psi(t)\rangle =\int_{t_0}^t \Xi(t-t^{\prime})\langle
\psi(t^{\prime})\rangle dt^{\prime} \equiv (\Xi\star\langle
\psi\rangle)(t)\,\,,\,\\
\langle\psi(t)\rangle =[\,1+(i/\hbar)\,G_0\star \,\Xi\,\star
\,\,]^{-\,1}\,\Psi^0(t,0)\, \label{mo}
\end{array}
\end{equation}
The kernel $\,\Xi(\tau)\,$ of the ``mass operator''
$\,\Xi\,\star\,\,$ looks as
\begin{equation}
\begin{array}{c}
\Xi(\tau)=\overline{x}\,\delta(\tau)-\frac {i}{\hbar}\,
G_x(\tau)\langle\,\xi(\tau),\xi(0)\rangle\,+\,...\,\,,
\end{array} \label{xiop}
\end{equation}
where dots substitute higher-order ``non-factorable Feynman
diagrams'', and
\begin{equation}
\begin{array}{c}
G_x(\tau)\equiv [1+(i\overline{x}/\hbar
)\,G_0\,\star\,\,]^{-1}G_0(\tau)
\end{array} \label{gx}
\end{equation}

Correspondingly, Eq.\ref{mre1} together with (\ref{mo})-(\ref{xiop})
yields
\begin{equation}
\begin{array}{c}
\langle\Re\rangle =\text{Re\,}
\,C/(1+C)\,\,,\,\,\,\langle\mathcal{T}\rangle =\text{Re\,}
\,1/(1+C)\,\,\,,\,\,\label{rr}\\
C\equiv (i/\hbar v_0)\int_0^{\infty}\,
\Xi(\tau)\exp(iE_0\tau/\hbar)\,d\tau\,\,\,\,\,\,\,
\end{array}
\end{equation}
We noticed that
$\,\int_0^{\infty}\,G_0(\tau)\exp(iE\tau/\hbar)\,d\tau =1/V(E)\,$,
where $\,V(E)=\sqrt{E(\Delta E -E)}/\hbar\,$ is particle's velocity
as function of its energy (thus $\,v_0=V(E_0)\,$).

Because of the stationarity, both $\,\langle\psi(t)\rangle\,$ and
$\,\langle\xi(t)\psi(t)\rangle\,$ oscillate like the source
$\,\Psi^0(t,0)\,$ in (\ref{ise1}): ~$\,\langle\psi(t)\rangle \propto
\langle\xi(t)\psi(t)\rangle \propto \exp(-iE_0 t/\hbar)\,$. Hence,
any inelastic amplitude of (\ref{amps}) is zero on average:
\begin{equation}
\begin{array}{c}
\langle A_k^S\rangle \propto \,\delta(|k|-|k_0|)\,\,\label{ma}
\end{array}
\end{equation}

From this point, concentrate on simplest but principal situation
characterized by three assumptions as follow.

({\bf i}) $\,\xi(t)$'s fluctuations are produced by equilibrium
thermal bath, therefore, according to (\ref{cf})-(\ref{kxy}),
\begin{equation}
\langle\,\xi(\tau),\xi(0)\rangle\,=K(|\tau |)\,\,,\,\,\,
\langle\,\xi^{*}(\tau),\xi(0)\rangle\,=K(\tau )\,\,,\label{kk}
\end{equation}
\begin{equation}
K(\tau)=\int_0^{\infty}\frac
{e^{i\omega\tau}+\exp(\hbar\omega/T)e^{-i\omega\tau}}
{1+\exp(\hbar\omega/T)}\,\sigma(\omega)\,d\omega\,\,,\label{kxi}
\end{equation}
where $\,T\,$ is bath temperature and $\,\sigma(\omega)\geq 0\,$. The
spectrum $\,\sigma(\omega)$ can be naturally introduced through
correlation of the random potential $\,x(t)=\text{Re\,}\xi(t)$:
~$\,\langle\,x(\tau),x(0)\rangle\,=\int_0^{\infty}\,\cos(\omega\tau)
\sigma(\omega)\,d\omega$.

({\bf ii}) $\,\xi(t)$'s fluctuations are sufficiently small and
short-correlated (wide-band), e.g. in the sense of
\begin{equation}
\begin{array}{c}
\max\,\sigma(\omega)\,\ll\,\hbar\Delta E\,\,,\label{cond}
\end{array}
\end{equation}
so that the scattering can by analyzed in the ``Born approximation''.

({\bf iii}) $\,\overline{x}=0\,$, i.e. ``there is no potential
irregularity and no scattering on average''.

Under these conditions, when averaging $\,\Re\,$ and other variables
one can manage with the correlation function $\,K(\tau)\,$ only, as
if $\,\xi(t)\,$ was Gaussian noise. Due to (\ref{cond}), formula
(\ref{rr}) strongly simplifies to
\begin{equation}
\langle\Re\rangle\,\approx \text{Re\,}C\,\approx \frac
{1}{v_0\hbar^2}\int_{-E_0/\hbar}^{(\Delta E-E_0)/\hbar}\,\frac
{\sigma(|\omega|)F(\hbar\omega/T)}{V(E_0+\hbar\omega)}
\,d\omega\,\,,\label{rc}
\end{equation}
where $\,F(X)\equiv 1/[1+\exp(X)]\,$ is the Fermi distribution
function, and presumably $\,\text{Re\,}C\,<1\,$. The contributions to
$\,C\,$ from $\,\omega
>0\,$ and $\,\omega <0\,$ correspond to scattering
with radiation or absorption of energy by the bath, respectively.

When $\,\overline{x}=0\,$, then elastic backward scattering must
disappear at $\,\sigma\rightarrow 0\,$, that is $\,\langle
A^S_{-k_0}\rangle\propto \sigma\,$ at small $\,\sigma\,$. What is for
fluctuational contribution, $\,\langle
A_{-k_0}\,,A_{-k_0}^{*}\rangle\,$, to mean probability of this
scattering, it is small to the extent of $\,\sigma/L\,$, as well as
$\,\langle A_k\,,A_k^{*}\rangle$ at all (see Eq.\ref{af} below).
Hence, $\,\langle\Re^{el}\rangle \ll \langle\Re\rangle\,$, and the
quantity (\ref{rc}) in fact represents the probability of inelastic
reflection $\,\langle\Re^{in}\rangle\,$.

The latter statement can be confirmed by direct calculation of
$\,\langle A_k\,,A_k^{*}\rangle\,$ using Eq.\ref{amps} and the
approximation
\[
\begin{array}{c}
\langle\xi^{*}(t^{\prime})\psi^{*}(t^{\prime})\,,
\xi(t)\psi(t)\rangle\approx
\langle\xi^{*}(t^{\prime})\,,\xi(t)\rangle\,\langle
\psi^{*}(t^{\prime})\rangle\,\langle\psi(t)\rangle\,,
\end{array}
\]
which is valid under above assumptions. The result is
\begin{equation}
\langle A_k\,,A_k^{*}\rangle = \frac {2\pi\, }{L\,v_0\hbar ^2
}\,\,\sigma (|\varepsilon (k)|/\hbar)\,F(\varepsilon (k)/T)\,\,,
 \label{af}
\end{equation}
where $\,\varepsilon (k)\equiv E(k)-E_0\,$. With use of the equality
(\ref{ma}), the inelastic reflection probability is
\begin{equation}
\begin{array}{c}
\langle\Re^{in}\rangle = \sum_{\,-k_0\,\neq\, k\,<\,0\,} \langle
A_k\,,A_k^{*}\rangle\,\, \label{rin}
\end{array}
\end{equation}
In fact, of course, for the finite incoming wave packet, this sum
contains a finite number $\,\approx L\,$ only of distinguishable
outgoing states. The factor $\,2\pi/L\,$ in (\ref{af}) is just their
separation in the $\,k$-space, and because of $\,L\gg 1\,$ the sum
(\ref{rin}) coincides with r.h.s of (\ref{rc}), that is
$\,\langle\Re^{in}\rangle\approx\langle\Re\rangle $.

{\bf\,VI.\,}~ We illustrated once again that the general stochastic
representation approach to open systems as suggested in \cite{i1} and
developed in \cite{i2,i3,i4,i5} can be useful in concrete
considerations replacing complicated dynamical evolution of composite
systems by stochastic evolution of a simple partial system. In the
above example, if the energy of incoming particle is not fixed but
has thermally equilibrium distribution with the same temperature as
the scatterer's temperature, then it is easy to verify that the same
distribution will be reproduced after the stochastic inelastic
scattering. Hence, we also obtain additional evidences that the
stochastic representation ensures automatical agreement with
thermodynamics.

To conclude, notice that our approach allows to set the problems
about transport processes and their noise (including fluctuations in
transport rates, e.g. in electrical conductivity), as well as about
dephasing of coherent quantum processes, in a new fashion, with
attraction of reach results on linear dynamic systems with randomly
varying parameters (in particular, on wave propagation in random
media).



\end{document}